\documentclass[doublecol]{epl2}

\title{Collective influence in evolutionary social dilemmas}
\shorttitle{Collective influence in evolutionary social dilemmas}

\author{Attila Szolnoki,\inst{1} Matja{\v z} Perc,\inst{2,3}}
\shortauthor{Szolnoki and Perc}
\institute{\inst{1}Institute of Technical Physics and Materials Science, Centre for Energy Research, Hungarian Academy of Sciences, P.O. Box 49, H-1525 Budapest, Hungary\\
\inst{2}Faculty of Natural Sciences and Mathematics, University of Maribor, Koro{\v s}ka cesta 160, SI-2000 Maribor, Slovenia\\
\inst{3}CAMTP -- Center for Applied Mathematics and Theoretical Physics, University of Maribor, Krekova 2, SI-2000 Maribor, Slovenia}


\abstract{When evolutionary games are contested in structured populations, the degree of each player in the network plays an important role. If they exist, hubs often determine the fate of the population in remarkable ways. Recent research based on optimal percolation in random networks has shown, however, that the degree is neither the sole nor the best predictor of influence in complex networks. Low-degree nodes may also be optimal influencers if they are hierarchically linked to hubs. Taking this into account leads to the formalism of collective influence in complex networks, which as we show here, has far-reaching implications for the favorable resolution of social dilemmas. In particular, there exists an optimal hierarchical depth for the determination of collective influence that we use to describe the potency of players for passing their strategies, which depends on the strength of the social dilemma. Interestingly, the degree, which corresponds to the baseline depth zero, is optimal only when the temptation to defect is small. Our research reveals that evolutionary success stories are related to spreading processes which are rooted in favorable hierarchical structures that extend beyond local neighborhoods.}

\begin{document}

\maketitle

The study of evolutionary games in structured populations has become a popular theme in statistical physics research, as evidenced by recent reviews that are devoted to this fascinating field \cite{szabo_pr07, roca_plr09, perc_bs10, rand_tcs13, perc_jrsi13, pacheco_plrev14, wang_z_epjb15}. Especially the evolution of cooperation in the realm of social dilemmas --- when individuals are torn between what is best for them and what is best for their society --- has received ample attention \cite{zimmermann_pre04, santos_pnas06, fu_pre09, du_wb_epl09, gao_srep15, xu_srep15,lee_s_prl11, gomez-gardenes_epl11, tanimoto_pre12, santos_md_srep14, pavlogiannis_srep15, wu_zx_epl15, hindersin_pcbi15, liu_rr_epl15, zimmermann_pre05, chen_w_pa16}. It has been shown that phase transitions leading to favorable evolutionary outcomes depend sensitively on the structure of the interaction network and the type of interactions, as well as on the number and type of competing strategies \cite{santos_prl05, gomez-gardenes_prl07, assaf_prl12, xu_b_srep15, javarone_csn15}.

Following the seminal discovery of Nowak and May that spatial structure can promote the evolution of cooperation \cite{nowak_n92b} through the mechanism now widely referred to as network reciprocity \cite{nowak_s06}, Santos and Pacheco where the first to show just how important the structure of the interaction network can be \cite{santos_prl05}. Players with the largest degree within the network --- the so-called hubs --- have the ability to dominate the evolution, which results in a strongly augmented network reciprocity in heterogeneous networks \cite{pacheco_ploscb06, szabo_pr07}. However, this positive effect can be destroyed quite easily if we take into consideration the realistic fact that more links with the neighbors also require more efforts to be maintained \cite{masuda_prsb07}. The simplest and most natural way to incorporate this into the traditional formalism of evolutionary games is to normalize the accumulated payoff by the number of links each player has. In doing so, hubs seize to matter, with the ultimate consequence being that the cooperation level drops to the comparatively low level that is characteristic for homogeneous networks and lattices \cite{tomassini_ijmpc07, szolnoki_pa08}.

But even if the positive effects of heterogeneous interaction networks are washed away by payoff normalization, a fact remains that our societies are not made up of uniform individuals. Rather contrary, inequalities abound, and not just among humans. We differ significantly not just in the number of friends and partners \cite{christakis_09}, but also in wealth, reputation, and influence that we are able to exert, to name but a few prominent examples. A very simple, minimalist way to model this observation is if we assume that players have unequal potency in passing a successful strategy to their neighbors during the strategy imitation process \cite{szolnoki_epl07}. It has been shown that the diversity of this potency --- sometimes referred to also as the teaching activity --- leads to a significantly increased cooperation level even if degree-normalized payoffs are used. In particular, if the teaching activity is proportional to the degree of a player then the originally high cooperation level that is characteristic for heterogeneous interaction networks is recovered. In this way social diversity can restore the elevated level of cooperation in a much broader context, and independently of the origin of heterogeneity or the applied social dilemma \cite{perc_pre08, yang_hx_pre09}.

Importantly, social diversity reintroduces the importance of different levels of influence in the evolutionary setting, the origin of which has recently been studied in the realm of social networks \cite{pinheiro_prl14}, and which also manifest as emergent hierarchical structures in evolutionary games \cite{lee_s_prl11}. In terms of the network perspective of influence, Morone and Makse have recently introduced the concept of collective influence \cite{morone_n15}, which pioneers a more wholesome take on the problem that goes beyond an individual's degree and hub status. In a nutshell, many low-degree nodes can be optimal influencers if they are ``surrounded by hierarchical coronas of hubs'', but they can only be uncovered through the optimal collective interplay of all
the influencers in the network. This arguably has far-reaching implications for a number of important social phenomena, ranging from efficient immunization \cite{pastor-satorras_rmp15} to the diffusion of information \cite{hidalgo_15}. As we will show in what follows, the collective influence plays an important role also in the evolution of cooperation in social dilemmas, in particular if a player's potency to pass its strategy maps to it.

To demonstrate our point, we use the simplified version of the prisoner's dilemma game,
where the key aspects of the social dilemma are preserved but its strength is determined by a single parameter \cite{nowak_n92b}. In particular, mutual cooperation yields the reward $R=1$, mutual defection leads to punishment $P=0$, and the mixed choice gives the cooperator the sucker's payoff $S=0$ and the defector the temptation $T>1$. We note, however, that the selection of this widely used and representative parametrization gives results that remain valid in a broad range of pairwise social dilemmas, including the snowdrift or the stag-hunt game.

We consider heterogeneous interaction networks, where each node is initially designated either as cooperator ($C$) or defector ($D$) with equal probability, and the evolutionary process in simulated in accordance with the standard Monte Carlo simulation procedure comprising the following elementary steps. First, according to the random sequential update protocol, a randomly selected player $x$ acquires its payoff $\Pi_x$ by playing the game with all its $k_x$ neighbors. Next, player $x$ randomly chooses one neighbor $y$, who then also acquires its payoff $\Pi_y$ in the same way as previously player $x$. Once both players acquire their payoffs, we normalize $\widetilde{\Pi}_x=\Pi_x/k_x$ and $\widetilde{\Pi}_y=\Pi_y/k_y$, thus taking into account the cost of maintaining the links and so avoid unrealistically high payoff values. Lastly, if $\widetilde{\Pi}_y > \widetilde{\Pi}_x$ then player $x$ adopts the strategy $s_y$ from player $y$ with the probability
\begin{equation}
\Gamma = w_y \frac{\widetilde{\Pi}_y - \widetilde{\Pi}_x}{T \cdot k_m}\,,
\label{prob}
\end{equation}
where $k_m$ is the largest of $k_x$ and $k_y$ \cite{santos_prl05}, and $0 < w_y \le 1$ determines the potency of player $y$ to pass its strategy \cite{szolnoki_epl07}.

Following \cite{morone_n15}, we determine the collective influence of a node $i$ as the product of its reduced degree and the total reduced degree of all $j$ nodes at a hierarchial depth $\ell$ from node $i$, according to
\begin{equation}
\textrm{CI}_\ell(i) = (k_i - 1) \sum_{d(i,j)=\ell} (k_j -1) \,.
\label{CIE}
\end{equation}
In order to use the collective influence to describe the potency of players for passing their strategies, we assume that
\begin{equation}
w_i = \frac{\textrm{CI}_\ell(i)}{\textrm{CI}_\ell}\,,
\label{map}
\end{equation}
where $\textrm{CI}_\ell(i)$ is the collective influence of node $i$ at depth $\ell$, while $\textrm{CI}_l$ is its maximum value among all nodes. The reference case, when the teaching activity is simply proportional to the degree of a node according to $w_i = k_i / k_{max}$, can be considered as the $\ell=0$ limit of this mapping. We emphasize that $\ell$ should be kept beyond a threshold value, as otherwise one can easily reach the diameter of the network at a fixed system size, as it is pointed out in \cite{morone_n15}.

For the main results presented below we use two representative heterogeneous networks, which are the scale-free network with the average degree $\langle k \rangle=4$ and the Erd{\H os}-R{\'e}nyi random network with $\langle k \rangle=12$ \cite{barabasi_s99}. By using networks consisting of $N=2 \cdot 10^5$ players, we are able to study the collective influence strength up to the $\ell=5$ hierarchical depth. We note that each full Monte Carlo step (MCS) consists of $N$ elementary steps described above, which are repeated consecutively, thus giving a chance to every player to change its strategy once on average. We determine the fraction of cooperators $f_C$ in the stationary state after a sufficiently long relaxation time lasting up to $10^6$ MCS. To further improve accuracy, the final results are averaged over $100$ independent realizations, including the generation of the interaction networks and random initial strategy distributions, for each set of parameter values.

\begin{figure}
\begin{center}
\includegraphics[width=8.5cm]{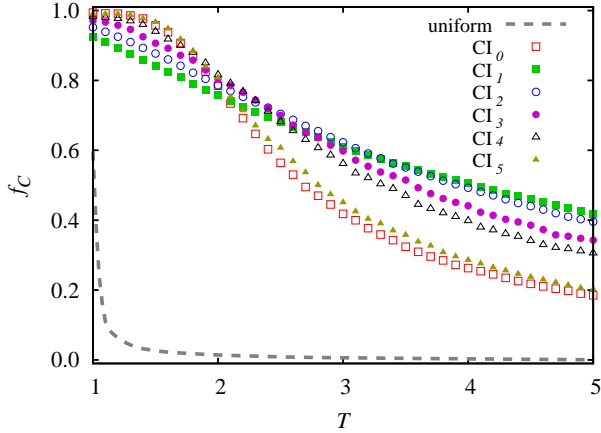}
\caption{\label{SF} Fraction of cooperators in dependence on the temptation to defect $T$, as obtained on scale-free networks where the strategy-pass potency of players is determined by their collective influence (CI). Indexes in the legend mark the hierarchical depth that was used to determine collective influence according to Eq.~\ref{CIE}. Here $0$ denotes the baseline case where CI is simply the degree of each player. For comparison, the outcome of the uniform model, where all players have $w_x=1$ is also shown. We note that in all cases degree-averaged payoffs were used in Eq.~\ref{prob}. It can be observed that, depending on the value of T, different hierarchical depths for determining CI yield the highest level of cooperation. For the most adverse conditions CI$_{\it 1}$ works best, which even at $T=4$ enables 50\% of the population to adopt cooperation.}
\end{center}
\end{figure}

We begin by showing the stationary fraction of cooperators in dependence on the temptation to defect on scale-free networks in Fig.~\ref{SF}. To illustrate the importance of diversity, we also show the results obtained with the uniform model, where all players have an identical potency to pass their strategy. In the latter case, due the application of degree-normalized payoffs, the presence of hubs gives no advantage to cooperators (as would otherwise be the case on a scale-free network \cite{santos_prl05}), and the cooperation level therefore falls immediately as we go beyond the $T=1$ value. When the diversity amongst players is restored, for example by making their teaching activity $w_i$ proportional to their degree $k_i$, then $f_C$ becomes high and decays only for large values of $T$. We refer to this as the baseline case, which is denoted as CI$_{\it 0}$ in Fig.~\ref{SF}. Interestingly, the application of collective influence to determine the strategy pass potency can improve the cooperation level further. Only up to $T=1.5$ CI$_{\it 0}$ provides the highest level of cooperation and can thus be considered optimal. For all higher values of $T$ the collective influence, determined at different hierarchical depths $\ell$ depending on the value of $T$, is a better proxy for the mapping of the potency of players. For example, at CI$_{\it 1}$ the majority of the population remains in a cooperative state up to $T=4$, which is twice the maximal value of $T$ that is traditionally considered as the upper limit for the prisoner's dilemma game. It can also be observed that, as we increase the hierarchical depth to determine the collective influence of players and accordingly their strategy-pass potency, the impact gradually decreases. For $\ell>4$ we practically recover the baseline CI$_{\it 0}$ case.

\begin{figure}
\begin{center}
\includegraphics[width=8.5cm]{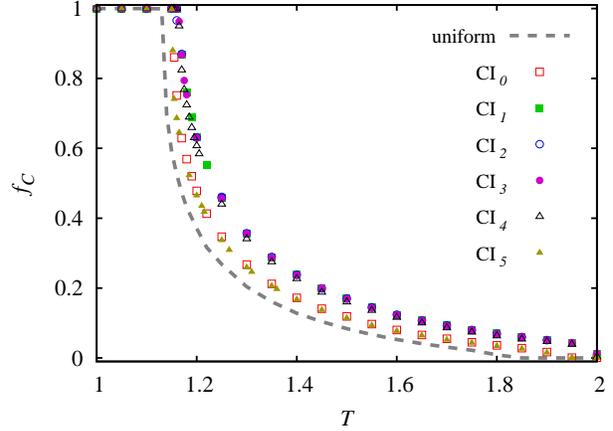}
\caption{\label{ER} Fraction of cooperators in dependence on the temptation to defect $T$, as obtained on Erd{\H o}s-R\'enyi random networks. As in Fig.~\ref{SF}, here too the strategy-pass potency of players is determined by their collective influence (CI), and indexes in the legend mark the hierarchical depth that was used to determine CI. The outcome of the uniform model is also shown. It can be observed that, largely regardless of the value of T, the highest level of cooperation is warranted by the CI$_{\it 1}$ case. The differences between different values of $\ell$ are not as well expressed as when scale-free networks were used, which has to do with the lower level of heterogeneity that characterizes random networks. Nevertheless, the CI$_{\it 0}$ case and the CI$_{\it 5}$ case are hardly distinguishable, marking the progressive worsening of the conditions for cooperation for $\ell>1$.}
\end{center}
\end{figure}

Results presented in Fig.~\ref{ER} show that all the relevant features related to the promotion of cooperation, which we have outlined when considering scale-free networks, remain intact also on Erd{\H o}s-R\'enyi random networks. Namely, the lowest cooperation level is obtained when all players are equal in their ability to pass strategies onto their neighbors. If the teaching activity $w_i$ is proportional to the degree $k_i$ (CI$_{\it 0}$ case), which here introduces binomially distributed diversity, the level of cooperation improves. Yet these levels can be improved even further if instead of simply the degree the collective influence determines the potency of players to pass their strategies. As in Fig.~\ref{SF}, the first hierarchical depth, yielding the CI$_{\it 1}$ case, ensures the biggest improvement. Subsequent applications of deeper levels to determine collective influence yield progressively worse outcomes, to the point that the CI$_{\it 0}$ case and the CI$_{\it 5}$ case are hardly distinguishable. As expected, however, the whole effect is less striking here due to the much more moderate heterogeneity of the interaction network.

\begin{figure}
\begin{center}
\includegraphics[width=8.5cm]{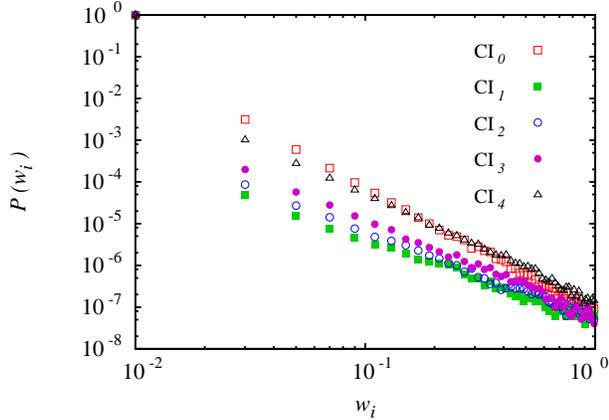}
\caption{\label{distSF} Distribution of strategy-pass potency of players, or teaching activities $w_i$, as obtained for different hierarchical depths $\ell$ that were used to determine the collective influence (CI) on scale-free networks. It can be observed that Eq.~\ref{map} maps the scale-free degree distribution faithfully onto the potency of players to pass their strategies, and this not only in the baseline case where CI is simply the degree of each player (which is naturally expected), but also for all higher depths $\ell$ that were used for determining CI. Interestingly, the slope of the power law distribution drops sharply when going from CI$_{\it 0}$ to CI$_{\it 1}$, only to then recover gradually to the baseline case for larger values of $\ell$.}
\end{center}
\end{figure}

To understand these results, in particular why the collective influence proposed by Morone and Makse \cite{morone_n15} yields more favorable outcomes than if simply the degree would be used as the base for determining the potency of players for passing their strategies, we explore the distributions of $w_i$ obtained for different values of hierarchical depth $\ell$. Figure~\ref{distSF} shows the results obtained for scale-free networks. The comparison reveals that the largest deviation from the original scale-free distribution is obtained for $\ell=1$.
In this case, the population practically segregates into two classes, such that there are some influential players with an intact strategy pass capacity, while the rest of the population is almost sterile. Conversely, the $P(w_i)$ function turns back to the baseline CI$_{\it 0}$ case gradually as we increase $\ell$. While the power law property is mostly preserved, the slope drops sharply when going from CI$_{\it 0}$ to CI$_{\it 1}$. This indicates that the application of collective influence for determining the teaching activity of players, in particular at $\ell=1$, significantly lowers the ability of ordinary, common individuals to pass their strategies. On the other hand, the most influential players, which have the highest CI, retain their high $w_i$ values.

\begin{figure}
\begin{center}
\includegraphics[width=8.5cm]{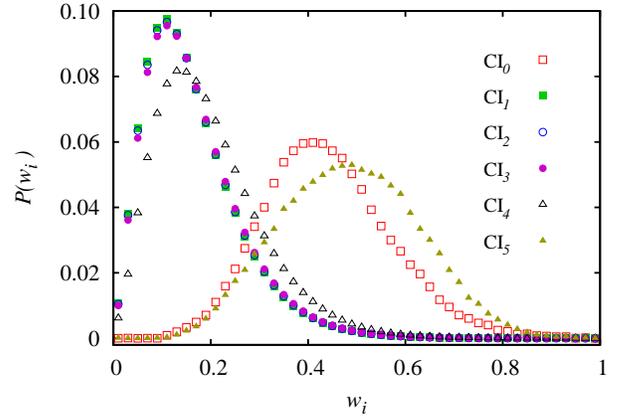}
\caption{\label{distER} Distribution of strategy-pass potency of players $w_i$, as obtained for different hierarchical depths $\ell$ that were used to determine the collective influence (CI) on Erd{\H o}s-R\'enyi random networks. Here again the properties of the binomial degree distribution of the network are reflected perfectly in the CI$_{\it 0}$ case, but are qualitatively preserved also for all higher hierarchical depths that were used to determine CI. As in Fig.~\ref{distSF}, the biggest difference in comparison to the CI$_{\it 0}$ case can be observed for $\ell=1$, when the peak of the bimodal distribution moves significantly to the left, thus indicating that the majority of players will have a much lower $w_i$. This difference subsequently disappears more and more for larger values of $\ell$.}
\end{center}
\end{figure}

Qualitatively identical results as shown in Fig.~\ref{distSF} are also obtained or Erd{\H o}s-R\'enyi random networks. Based on the results presented in Fig.~\ref{distER}, it can be observed that, starting from a binomial distribution at $\ell=0$, the biggest shift occurs for $\ell=1$, where the average strategy-pass potency decreases with the application of collective influence. When higher values of $\ell$ are applied, however, the baseline distribution is practically completely recovered, much in the same way as is the stationary fraction of cooperators in dependence on $T$ depicted in Fig.~\ref{ER}.

The presented distributions of teaching activity emphasize that those players who have the highest collective influence are in a very special situation because their potency to spread successful strategies is much better and in fact more critical than that of all the other players. This is true even if compared with the distributions that we have obtained with purely degree-related teaching activities on scale-free networks. In the latter case, hubs are frequently surrounded by other players who also have a high degree, and thus a high $w_i$ value. Our observations, combined with those reported by Morone and Makse, can explain the positive impact of CI-related teaching activities. As pointed out in \cite{morone_n15}, players with a high collective influence can bridge smaller communities that would reside in fragmented sub-domains in the lack of influencers. Accordingly, if we remove the nodes having high collective influence then the whole network would fall to pieces. Since the strategy-pass potency of players in the smaller, potentially isolated, domains is reduced, this is beneficial for cooperators. The asymmetric impact of weakened strategy invasion capabilities was already observed in \cite{szolnoki_pre09}, and it is based on the simple fact that defectors cannot invade their neighbors so easily, and hence cannot invade new ground. Strategies are thus faced with the consequences of their nature more frequently, which clearly supports network reciprocity because cooperators enjoy the neighborhood of akin players. Moreover, slow, coordinated invasions are always better for cooperators than they are for defectors, as it was demonstrated directly in \cite{szolnoki_njp13}.

In addition to the fact that the chance of a successful cooperation invasion is enhanced in smaller domains, there is another reason why the application of collective influence is better than the application of degree. Namely, players who are in ``gateway positions'' typically still have a high strategy pass capacity, and hence can exchange information between homogeneous smaller clusters efficiently. In this way, a mechanism similar to multilevel selection emerges between homogeneous clusters of different strategies, in that these clusters effectively act as players with the gatekeepers pushing their strategies on to other clusters.

\begin{figure}
\begin{center}
\includegraphics[width=8.5cm]{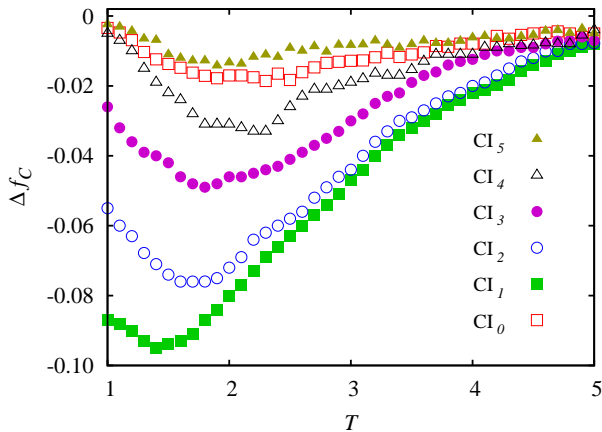}
\caption{\label{decay} Decay of the cooperation level on scale-free networks when top 2 $\%$ of players with the highest potency to pass strategy are sterilized and so made unable to be strategy donors. It can be observed that the population is most vulnerable when the strategy-pass potencies of players $w_i$ is determined according to the collective influence at hierarchial depth $\ell=1$. Importantly, this is also the depth at which cooperation fares best under the most adverse conditions (see Fig.~\ref{SF} for details). The baseline CI$_{\it 0}$ case and CI$_{\it 5}$ using the widest neighborhood to determine collective influence are affected least.}
\end{center}
\end{figure}

A straightforward consequence of the presented argument is that players with a high collective influence play a key role not only when it comes to effective immunization or diffusion of information, or other dynamical processes taking place on complex networks \cite{morone_n15, hu_prx14}, but also when it comes to the evolution of cooperation in that the outcome of the entire evolutionary processes is very sensitive to their state. This point can be easily verified simply by manipulating the players with the highest collective influence --- for example, by making them sterile. More precisely, we disable the top 2\% of players by denying them the ability to pass their strategy onto to their neighbors. For clarity, we restrict ourselves by showing the consequences on the scale-free network only, because there the difference between the top influencers and commoners is significantly more striking than on random networks.

In Fig.~\ref{decay}, we show how the cooperation level changes in dependence on the temptation $T$ due to the sterilization of the top 2\%. When $w_i$ is related to $k_i$, which corresponds to the CI$_{\it 0}$ case, then blocking the main hubs will not significantly lower the cooperation level in the population. On the one hand, this is because the distribution of $w_i$ remains strongly heterogeneous given that only part of the tail of the scale-free distribution is removed. On the other hand, players in the neighborhoods of hubs can replace the role of the latter very effectively. In particular, they can still transfer the more successful strategy and hence keep the selection process intact. But when the potency to pass a strategy is related to the collective influence, then the population becomes much more vulnerable to the loss of the top 2\%. In the latter case a more obvious decay of cooperation level can be observed in Fig.~\ref{decay}. There are two main reasons for this difference. In the first place, blocking the main influencers will result in a less heterogeneous population because all the other players have a significantly lower potency to pass strategy (this is illustrated in Fig.~\ref{distSF}). Secondly, and more importantly, sterilizing CI leaders will block the information flow across the main gateway positions, and thus hinder the previously intact emergence of multilevel selection between the smaller domains. While these two factors exist for all $\ell \geq 0$, they are the most efficient when the depth of influence is $\ell=1$. We note that the loss of cooperation expectedly becomes weaker at very high $T$ values because then there is no real competition between cooperators and defectors.

In summary, we have studied the role of collective influence in evolutionary social dilemmas, showing that there exists an optimal hierarchical depth for the determination of this influence that favors cooperation. While for low temptations to defect simply the degree, which corresponds to depth zero, works fine and is in fact optimal, for larger temptations collective influence based on depths one or two is best, depending slightly also on the structure of the interaction network.

We have shown that the players who have high collective influence are able to optimally transfer their strategies to others in the population. As Morone and Makse have pointed out \cite{morone_n15}, nodes with high collective influence can bridge smaller communities, and their removal can cause the whole network to disintegrate. Because we have used collective influence to describe the potency of players for passing their strategies, this effectively means that the potency of players in the smaller domains is reduced. In turn, this is beneficial for cooperators because of the asymmetric impact of weakened strategy invasion capabilities, which we have observed before in \cite{szolnoki_pre09}. More generally, the lessened potency of players in the smaller domains also slows down the evolutionary process, which always favors cooperators more than it does defectors, as we have demonstrated directly in \cite{szolnoki_njp13}.

In addition to the favorable adjustments of strategy invasions, the implications of collective influence spell benefits also at the so-called ``gateway positions'', where players still have a high potency to pass their strategies. We have argued that this leads to the emergence of multilevel selection involving smaller homogeneous domains of different strategies. Similar positive effects of spontaneously induced multilevel selection have been reported before in the realm of coevolving random networks \cite{szolnoki_njp09}, and conceptually similar effects were also detected in the realm of evolutionary games on minimally structured populations \cite{szollosi_pre08}. Moreover, such mechanisms can emerge even on regular graphs and lattices if the leading players are able to exchange information directly \cite{perc_pre08b}.

Our research highlights the continuation of the fruitful interplay between evolutionary game theory and network science \cite{szabo_pr07, perc_bs10}. A decade ago it was discovered that the diversity of players could have a decisive role for the successful evolution of cooperation in social dilemmas \cite{santos_pnas06}. Now we know that not merely the local properties, like the degree of each player, determine the evolutionary success stories, but that the sensitivity of the spreading process or the dynamics in the complex interacting network could also be decisive factors \cite{yang_hx_njp14, pei_jsm13}. Accordingly, hierarchical structures that extend beyond local neighborhoods must be taken into account.

\begin{acknowledgments}
This research was supported by the Hungarian National Research Fund (Grant K-101490) and the Slovenian Research Agency (Grants P5-0027 and J1-7009).
\end{acknowledgments}


\begin{thebibliography}{10}
\expandafter\ifx\csname url\endcsname\relax\def\url#1{\texttt{#1}}\fi

\bibitem{szabo_pr07}
\Name{Szab{\'o} G. \and F{\'a}th G.} \REVIEW{Phys. Rep.}{446}{2007}{97}.

\bibitem{roca_plr09}
\Name{Roca C.~P., Cuesta J.~A. \and S{\'a}nchez A.} \REVIEW{Phys. Life Rev.}{6}{2009}{208}.

\bibitem{perc_bs10}
\Name{Perc M. \and Szolnoki A.} \REVIEW{BioSystems}{99}{2010}{109}.

\bibitem{rand_tcs13}
\Name{Rand D.~A. \and Nowak M.~A.} \REVIEW{Trends in Cognitive Sciences}{17}{2013}{413}.

\bibitem{perc_jrsi13}
\Name{Perc M., G{\'o}mez-Garde{\~n}es J., Szolnoki A. \and Flor{\'{\i}a and Y.
  Moreno} L.~M.} \REVIEW{J. R. Soc. Interface}{10}{2013}{20120997}.

\bibitem{pacheco_plrev14}
\Name{Pacheco J.~M., Vasconcelos V.~V. \and Santos F.~C.} \REVIEW{Physics of
  Life Reviews}{11}{2014}{573}.

\bibitem{wang_z_epjb15}
\Name{Wang Z., Wang L., Szolnoki A. \and Perc M.} \REVIEW{Eur. Phys. J. B}{88}{2015}{124}.

\bibitem{zimmermann_pre04}
\Name{Zimmermann M.~G., Egu{\'{\i}}luz V.~M. \and San~Miguel M.} \REVIEW{Phys.
  Rev. E}{69}{2004}{065102(R)}.

\bibitem{santos_pnas06}
\Name{Santos F.~C., Pacheco J.~M. \and Lenaerts T.} \REVIEW{Proc. Natl. Acad.
  Sci. USA}{103}{2006}{3490}.

\bibitem{fu_pre09}
\Name{Fu F., Wu T. \and Wang L.} \REVIEW{Phys. Rev. E}{79}{2009}{036101}.

\bibitem{du_wb_epl09}
\Name{Du W.-B., Cao X.-B., Hu M.-B. \and Wang W.-X.} \REVIEW{EPL}{87}{2009}{60004}.

\bibitem{gao_srep15}
\Name{Gao L., Wang Z., Pansini R., Li Y.-T. \and Wang R.-W.}
\REVIEW{Sci. Rep.}{5}{2015}{17752}.

\bibitem{xu_srep15}
\Name{Xu B. \and Wang J.}
\REVIEW{Sci. Rep.}{5}{2015}{16447}.

\bibitem{lee_s_prl11}
\Name{Lee S., Holme P. \and Wu Z.-X.} \REVIEW{Phys. Rev. Lett.}{106}{2011}{028702}.

\bibitem{gomez-gardenes_epl11}
\Name{G{\'o}mez-Garde{\~n}es J., Vilone D. \and S{\'a}nchez A.} \REVIEW{EPL}{95}{2011}{68003}.

\bibitem{tanimoto_pre12}
\Name{Tanimoto J., Brede M. \and Yamauchi A.} \REVIEW{Phys. Rev. E}{85}{2012}{032101}.

\bibitem{santos_md_srep14}
\Name{Santos M., Dorogovtsev S.~N. \and Mendes J. F.~F.} \REVIEW{Sci. Rep.}{4}{2014}{4436}.

\bibitem{pavlogiannis_srep15}
\Name{Pavlogiannis A., Chatterjee K., Adlam B. \and Nowak M.~A.} \REVIEW{Sci.
  Rep.}{5}{2015}{17147}.

\bibitem{wu_zx_epl15}
\Name{Wu Z.-X., Rong Z. \and Chen M. Z.~Q.} \REVIEW{EPL }{110}{2015}{30002}.

\bibitem{hindersin_pcbi15}
\Name{Hindersin L. \and Traulsen A.} \REVIEW{PLoS Comput. Biol.}{11}{2015}{e1004437}.

\bibitem{liu_rr_epl15}
\Name{Liu R.-R., Jia C.-X. \and Rong Z.} \REVIEW{EPL}{112}{2015}{48005}.

\bibitem{zimmermann_pre05}
\Name{Zimmermann M.~G. \and Egu{\'{\i}}luz V.~M.} \REVIEW{Phys.
  Rev. E}{72}{2005}{056118}.

\bibitem{chen_w_pa16}
\Name{Chen W., Wu T., Li Z., \and Wang L.}
\REVIEW{Physica A}{443}{2016}{192}.

\bibitem{santos_prl05}
\Name{Santos F.~C. \and Pacheco J.~M.} \REVIEW{Phys. Rev. Lett.}{95}{2005}{098104}.

\bibitem{gomez-gardenes_prl07}
\Name{G{\'o}mez-Garde{\~n}es J., Campillo M., Flor{\'{\i}}a L.~M. \and Moreno
  Y.} \REVIEW{Phys. Rev. Lett.}{98}{2007}{108103}.

\bibitem{assaf_prl12}
\Name{Assaf M. \and Mobilia M.} \REVIEW{Phys. Rev. Lett.}{109}{2012}{188701}.

\bibitem{xu_b_srep15}
\Name{Xu B. \and Wang J.} \REVIEW{Sci. Rep.}{5}{2015}{16447}.

\bibitem{javarone_csn15}
\Name{Javarone M.~A. \and Atzeni A.~E.} \REVIEW{Comput. Soc. Netw.}{2}{2015}{15}.

\bibitem{nowak_n92b}
\Name{Nowak M.~A. \and May R.~M.} \REVIEW{Nature}{359}{1992}{826}.

\bibitem{nowak_s06}
\Name{Nowak M.~A.} \REVIEW{Science}{314}{2006}{1560}.

\bibitem{pacheco_ploscb06}
\Name{Pacheco J.~M., Santos F.~C. \and Chalub A. C.~C.} \REVIEW{PLoS Comp.
  Biol.}{2}{2006}{1634}.

\bibitem{masuda_prsb07}
\Name{Masuda N.} \REVIEW{Proc. R. Soc. B}{274}{2007}{1815}.

\bibitem{tomassini_ijmpc07}
\Name{Tomassini M., Luthi L. \and Pestelacci E.} \REVIEW{Int. J. Mod. Phys. C}{18}{2007}{1173}.

\bibitem{szolnoki_pa08}
\Name{Szolnoki A., Perc M. \and Danku Z.} \REVIEW{Physica A}{387}{2008}{2075}.

\bibitem{christakis_09}
\Name{Christakis N.~A. \and Fowler J.~H.} \Book{Connected: The Surprising Power
  of Our Social Networks and How They Shape Our Lives} (Little Brown, New York)
  2009.

\bibitem{szolnoki_epl07}
\Name{Szolnoki A. \and Szab{\'o} G.} \REVIEW{EPL}{77}{2007}{30004}.

\bibitem{perc_pre08}
\Name{Perc M. \and Szolnoki A.} \REVIEW{Phys. Rev. E}{77}{2008}{011904}.

\bibitem{yang_hx_pre09}
\Name{Yang H.-X., Wang W.-X., Wu Z.-X., Lai Y.-C. \and Wang B.-H.}
  \REVIEW{Phys. Rev. E}{79}{2009}{056107}.

\bibitem{pinheiro_prl14}
\Name{Pinheiro F., Santos M.~D., Santos F. \and Pacheco J.} \REVIEW{Phys. Rev.
  Lett.}{112}{2014}{098702}.

\bibitem{morone_n15}
\Name{Morone F. \and Makse H.~A.} \REVIEW{Nature}{524}{2015}{65}.

\bibitem{pastor-satorras_rmp15}
\Name{Pastor-Satorras R., Castellano C., Van~Mieghem P. \and Vespignani A.}
  \REVIEW{Rev. Mod. Phys.}{87}{2015}{925}.

\bibitem{hidalgo_15}
\Name{Hidalgo C.} \Book{Why Information Grows: The Evolution of Order, from
  Atoms to Economies} (Basic Books, New York) 2015.

\bibitem{barabasi_s99}
\Name{Barab{\'a}si A.-L. \and Albert R.} \REVIEW{Science}{286}{1999}{509}.

\bibitem{hu_prx14}
\Name{Hu Y., Havlin S. \and Makse H.~A.}
\REVIEW{Phys. Rev. X}{4}{2014}{021031}.

\bibitem{szolnoki_pre09}
\Name{Szolnoki A., Perc M., Szab{\'o} G. \and Stark H.-U.} \REVIEW{Phys. Rev. E}{80}{2009}{021901}.

\bibitem{szolnoki_njp13}
\Name{Szolnoki A. \and Perc M.} \REVIEW{New J. Phys.}{15}{2013}{053010}.

\bibitem{szolnoki_njp09}
\Name{Szolnoki A. \and Perc M.} \REVIEW{New J. Phys.}{11}{2009}{093033}.

\bibitem{szollosi_pre08}
\Name{Sz\"oll\H{o}si G.~J. \and Der{\'e}nyi I.} \REVIEW{Phys. Rev. E}{78}{2008}{031919}.

\bibitem{perc_pre08b}
\Name{Perc M., Szolnoki A. \and Szab{\'o} G.} \REVIEW{Phys. Rev. E}{78}{2008}{066101}.

\bibitem{yang_hx_njp14}
\Name{Yang H.-X., Rong Z. \and Wang W.-X.} \REVIEW{New J. Phys.}{16}{2014}{013010}.

\bibitem{pei_jsm13}
\Name{Pei S. \and Makse H.~A.}
\REVIEW{J. Stat. Mech.}{2013}{2013}{P12002}

\end{thebibliography}
\end{document}